# Enhanced Eshelby Twist on Thin Wurtzite InP Nanowires and Measurement of Local Crystal Rotation

L. H. G. Tizei,[1,2,*] A. J. Craven,[3] L. F. Zagonel,[2,4] M. Tencé,[2] O. Stéphan,[2]
T. Chiaramonte,[1] M. A. Cotta,[1] and D. Ugarte[1,†]

[1]Instituto de Física "Gleb Wataghin", Universidade Estadual de Campinas-
UNICAMP, Rua Sérgio Buarque de Holanda 777, 13083-859 Campinas, SP, Brazil
[2]Laboratoire de Physique des Solides, Université Paris-Sud, CNRS-UMR 8502, Orsay 91405, France
[3]SUPA School of Physics & Astronomy, University of Glasgow, Glasgow G12 8QQ, United Kingdom
[4]Laboratório Nacional de Nanotecnologia, C. P. 6192, 13083-970 Campinas, SP, Brazil

We have performed a detailed study of the lattice distortions of InP wurtzite nanowires containing an axial screw dislocation. Eshelby predicted that this kind of system should show a crystal rotation due to the dislocation induced torque. We have measured the twisting rate and the dislocation Burgers vector on individual wires, revealing that nanowires with a 10-nm radius have a twist up to 100% larger than estimated from elasticity theory. The strain induced by the deformation has a Mexican-hat-like geometry, which may create a tube-like potential well for carriers.

The new electro-optical properties of semiconductor nanowires (NWs) place them among promising nano-objects for future devices [1–4]. Although it is well known that strain effects can greatly influence semiconductor physical behavior [5–7], precise characterization of structural modifications in nanoscale materials is still an open and challenging problem, hindering the establishment of a basic understanding and development of NW technological applications. For example, in 1953, Eshelby predicted that a whisker containing a screw dislocation should show a crystal rotation due to the induced torque [8,9]. The Eshelby twist model, based on macroscopic elastic theory [8,9], predicts a rotation per unit length that depends on the magnitude of the Burgers vector $|b|$ [10] and on the cross sectional area, A, of the whisker with $\alpha|b|=A$. Recently [11,12], the Eshelby twist [8] has been observed for PbS and PbSe branched NWs containing an axial screw dislocation with a diameter in the 60–100 nm range. For PbSe, the estimation of the Burgers vector magnitude using the Eshelby model suggest an anomalous vector distribution, which the authors interpreted as an indication of super screw dislocations [12]. These kinds of dislocations have been shown to exist in open core systems [13–15]. However, their occurrence in solid core systems is unlikely due to the additional strain that must be accommodated. The inference of structural parameters of nanoscale objects using macroscopic models should be always corroborated by independent measurements. Briefly, measurements of $|b|$ are essential to understand the actual correlation between NWs twisting and screw dislocation properties. Although the determination of $|b|$ has been reported in defective GaN films [16], this method cannot be easily applied to thin nanowires. We have used an alternate technique based on electron diffraction to determine $|b|$. By measuring $|b|$ and in different NWs, we are able to report a larger (up to 100%) twist rate than predicted by the macroscopic model, revealing a change in physical behavior. We interpret this enhancement as caused by surface effects.

We have studied InP wurtzite NWs (radius in the 10–20 nm range), some of which contain a screw-type dislocation and display the Eshelby twist effect. These NWs have been grown by chemical beam epitaxy on nominal [001] GaAs substrates using the vapor-liquid-solid (VLS) mechanism [17,18] at 480°C for 45 min. Physically deposited Au nanoparticles have been used as catalysts [19]. The described NWs have been generated in growth experiments attempting to synthesize InGaP NWs. Although triethylgallium was introduced in the chamber, energy dispersive x-ray spectroscopy in a transmission electron microscopy (TEM) showed that the Ga content in thin NW regions is below the detection limit (about 5% in this case). Then, all wires analyzed here will be considered as pure InP. However, we do not believe that this fact is crucial to the conclusions of this work. The presence of Ga in the growth chamber may be related to the origin of the screw dislocation observed in some NWs and will be discussed in an ongoing work.

Transmission electron microscopy and atomic resolution TEM (HRTEM) were performed with a JEOL 2010. Scanning transmission electron microscopy (STEM) experiments were performed with a VG HB 501 operated at 60 kV and 100 kV. This instrument was used to acquire bright field (BF) STEM images as well as convergent beam electron diffraction (CBED). Scanning EM images have been acquired on a Gemini Zeiss microscope. Atomically resolved HAADF (high angle annular dark field) STEM images have been acquired on a Nion UltraSTEM 100 microscope operated at 60 kV.

Scanning EM images (see the Supplemental Material [20], Fig. S1) reveal two populations of NWs on the growth substrate: short ($< 2$ μm) and long ones ($> 4$ μm). TEM, HRTEM [Fig. 1(a)], and BF STEM [Fig. 1(b)] imaging show that longer NWs contain axial screw dislocations while shorter ones do not. This correlation between long NWs and the presence of a line defect suggests the existence of a screw dislocation assisted VLS growth, occurring simultaneously with the defect-free shorter NWs conventional VLS growth in the same sample. So far, screw dislocation-driven models have been proposed for the growth without catalysts [11,15,21,22]. In our case, however, the faster VLS growth in the presence of the defect is in agreement both with Burton-Cabrera-Franck's theory of crystal growth [23] and with earlier experimental observations for liquid phase epitaxy assisted by defects [24]. Our TEM samples have been prepared by mechanical transfer [20], showing that the dislocations are stable in NWs with 10 nm radii.

For the longer NWs [Fig. 1(b)], a progressive rotation of the atomic arrangement (twist) can be observed acquiring a sequence of CBED patterns along the NW axis (see [20], movie M1). This kind of NW shows [Fig. 1(c)] a dark central line that runs along the NW corresponding to the screw dislocation core and a periodic distribution of dark bands perpendicular to the axis. These bands can be separated in two subgroups: (a) primary dark bands, which are perpendicular to the NW growth direction; and (b) secondary side bands, which are not perpendicular to the growth direction and are discontinuous at the NW center ([20], Fig. S2). To understand this contrast, we have performed spatially resolved CBED measurements parallel and perpendicular to NWs' growth direction. These show that primary dark bands occur at regions close to a zone axis. Figures 1(d) and 1(e) show the experimental CBED patterns with the beam positioned on two neighboring primary bands. The CBED patterns' orientations can be identified as the [01$\bar{1}$0] and [2$\bar{1}\bar{1}$0] zone axes of the wurtzite structure, which means an axial NW rotation of 30° (524 mrad) over a 210 nm long segment or $(0.14\pm0.01)°$/nm [$(2.5\pm0.2)$ mrad/nm] twist. Finally, we have also analyzed in detail the secondary side bands, which shift abruptly as they cross the NW center. All features can be explained considering diffraction effects induced by the geometrical arrangement of atoms in a screw dislocation, in which atomic planes form a helicoid. In fact, the normal to the atomic planes is slightly tilted in relation to the twisting axis ([20], Fig. S3). The angle of tilt ($\theta$) between planes on opposite sides of the dislocation is related to the pitch of the helicoid, which is equal to $|b|$, and the radial position (a specific value of R) of the measurement point [Eq. (1)];

$$\cos(\theta) = \frac{1 - (\frac{b}{2\pi R})^2}{1 + (\frac{b}{2\pi R})^2}. \quad (1)$$

We must note that this tilt is in a direction perpendicular to the helicoid twist axis, making it impossible to mix up the two effects. Considering this geometry, the observed features in BF images are explained.

The direct relation between BF STEM image contrast bands and crystal orientation allows the quantification of the atomic structure twisting rate. Figure 2 shows 87 measured rates in 19 NWs as a function of the inverse of the cross sectional area. Using this plot and assuming the validity of Eshelby's model [8], one can calculate $|b|$ for the dislocation. The histogram of the estimated $|b|$ shows a peak at $(1.0\pm0.1)$ nm ([20], Fig. S4).

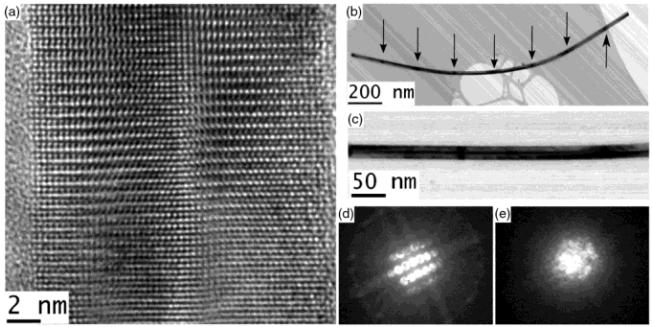

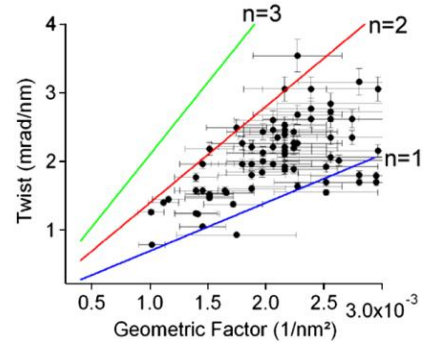

FIG. 1: (a) HRTEM image showing a discontinuity of the atomic planes at the center of the NW. The contrast changes along the NW growth direction. (b) BF STEM image of a NW showing a periodic set of dark bands (marked by arrows). (c) BF STEM image of the region from which the CBED movie M1 in the Supplemental Material [20] has been made. (d) and (e) Experimental CBED patterns from the two primary bands on c).

FIG. 2: NW twist rate as a function of the inverse of the cross sectional area. The blue, red, and green lines represent the twist rate for screw dislocations with $|b| = 0.68$ nm (n = 1), $|b| = 1.36$ nm (n = 2), and $|b| = 2.04$ nm (n = 3), respectively. Even considering the uncertainties on the cross sectional area, the twist data are not consistent with any of the suggested burgers vectors.

A wurtzite crystal can be seen as an *ABAB* stacking of atomic planes (each plane contains In-P dimers). This results in a hexagonal lattice with parameters $a = 0.47$ nm and $c = 0.68$ nm, along the basal plane and the $z$ axis, respectively. A and B planes are stacked in the $z$ direction but they present an $xy$ displacement. In a screw dislocation, $b$ is the displacement of atomic planes when a closed circuit around the dislocation line is completed. We have detected screw dislocations only along the [0001] direction. Then, we must consider that an A-type plane must connect to another A plane, as only z displacements are possible. Otherwise, additional $xy$ lattice shift components are needed at the *A-B* junctions, which were not detected. This geometrical constrain implies that $|b|$ must be an integer multiple of the *A-A* spacing (c); thus, $|b| = n*c$, with $n$ an integer.

HAADF images acquired using an aberration corrected STEM ([20], Fig. S5) indicate that the NWs are not hollow, excluding effects of an inner surface in strain relaxation. As the total energy of a dislocation scales with $|b|^2$, large n vectors should be less favorable in the solid core NWs studied here. With this in mind, we have compared in Fig. 2 the measured twist values with the predicted ones (lines, calculated using Eshelby model), when the Burgers vector correspond to n = 1, 2, or 3 (or $|b| = 0.68$, 1.36, and 2.04 nm). The experimental points fall clearly between the lines representing n = 1 and n = 2. At first sight, the effect of different cross sectional geometries [9] could be playing a role. However, the geometric prefactor correction κ ($α = κ|b|/A$), introduced by Eshelby [9], is 1.015 for a hexagonal cross section (typical for III-V semiconductor wurtzite nanowires), a 1.5% difference to the circular case. Even considering the difference in the total area (due to possible projection issues), the deviation in the calculated twist rate would be, at most, 20%, which is much smaller than the experimentally measured ones (up to 100%). At this point, to correctly interpret and discuss the measured twist values and the Eshelby model, it is essential to measure the Burgers vector magnitude.

Selected area electron diffraction (SAED) patterns [Fig. 3(a)] show a splitting of diffraction spots into two disks. A splitting of all spots by a fixed angle is expected for helicoidal structures because the atomic plane normal is slightly tilted away from the NW axis; this has been observed, for example, in DNA or chiral nanotubes [25,26]. The SAED pattern in Fig. 3(b) shows a value of the splitting $θ_{SAED}$ equal to $(2.0 ± 0.5)°$ splitting, which can be compared to Eq. (1).

A SAED measurement considers a long NW region (hundreds of nanometers) and averages the tilt over all radial positions (R). As the number of diffracting atoms increases with the radial distance, the average splitting angle must be weighted by R. The result of this weighted average for a $(12.0±0.5)°$ nm NW is 1.96° for $|b| = 0.68$ nm (n = 1), 4.07 for $|b| = 1.36$ nm (n = 2), and 6.07° for $|b| = 2.04$ nm (n = 3). The measured value of $θ_{SAED}$ above agrees with n = 1. The same result has been obtained in four different NWs.

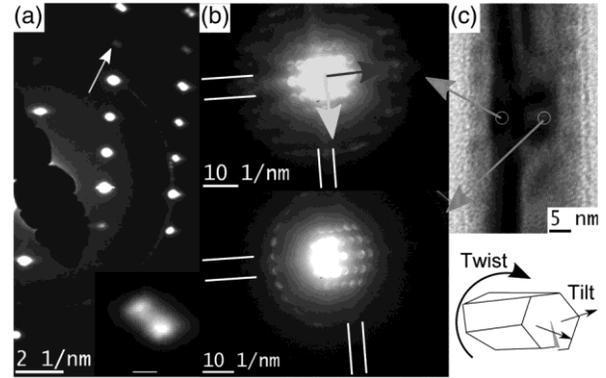

FIG. 3: (a) Selected area electron diffraction showing the splitting of the diffraction spots (the inset shows the split spot indicated by the arrow, scale 0.21/nm). (b) CBED patterns showing the tilt of the structure from the left (above) to the right part (below) of the section of the NW. (c) BF image showing the positions of measurement (grey circles) for each CBED pattern. The dark grey arrow points the twist axis while the light grey points the tilt axis in (b). This is shown schematically on the inset on the lower right. The Kikuchi bands observed have been used to measure the tilt angle.

The tilt of atomic planes can also be quantified by local crystallographic measurements using CBED. We have acquired CBED patterns in regular arrays, which we have called CBED imaging. The diffraction pattern changes abruptly when measured at different sides of the NW axis [Fig. 3(b)]. A detailed description of the tilt angle (θ) measurement using shifts in Kikuchi bands [10] is presented in [20], Fig. S6. The NWs studied here are thin. For this reason Kikuchi bands are faint; thus, we have performed these experiments using the lowest practical electron beam energy (60 kV). We must emphasize that the use of CBED allows the precise measurement of the tilt angle (error approximately 0.4°) for a defined R value. In these terms, we can invert Eq. (1) and derive $|b|$ from the tilt and R values. The average over 6 measurements in 4 different NWs (a fraction of the ones from which we have measured the twist rate) has yielded $|b| = (0.7 ± 0.1)$ nm.

HRTEM images show that the crystalline structure is strained, with the projection of the double-atomic planes showing the effect of them being in helix-like curves [Fig. 1(a)]. This agrees with the atomic arrangement of a screw dislocation, in which atomic planes form helicoids ([20], Fig. S3). Simulations with $|b| = 1.36$ nm (n = 2) and with $|b| = 2.04$ nm (n = 3) show intensity patterns (atomic planes tilting) which are more pronounced than those observed in experimental images ([20], Fig. S7. Simulations have been performed with the software QSTEM [27]). The experimental HRTEM images are consistent with the contrast pattern generated by a screw dislocation with $|b| = 0.68$ nm ($n = 1$). Finally, the contrast changes along the NW axis due to the twisting effect. The changes observed are consistent with a small twisting of the crystal lattice ([20], Fig. S7).

CBED and SAED experiments have provided a measurement of $|b|$ (in two different microscopes, a STEM and a TEM, and in different length scales), while HRTEM experiments have provided corroboration for this value. The ensemble of results agrees with $|b| = 0.68$ nm ($n = 1$), the expected one from crystallographic and energetic arguments. This result has been obtained without any hypothesis, except that the shape of a dislocated crystal is a helicoid, and by using measured parameters (and R). At this point, we have experimentally obtained all parameters involved in the Eshelby model ($\alpha$, NW radius and $|b|$). Given that $|b| = 0.68$ nm, the points in Fig. 2 should cluster around the line $n = 1$. With that, we can conclude that the plot in Fig. 2 indicates that the measured twist is larger than the predicted one and it reveals the occurrence of an enhanced twist in the InP NWs with approximately 10 nm radius.

We should point out that previous studies on NWs (PbS, [11], PbSe, [12]) dealt with NW radii which are much larger than reported here. As the proportionality constant between $|b|$ and the twist is the NW cross section, these NWs present a twist rate at least 1 order of magnitude smaller. Our results suggest that size effects must be playing a dominant role; as the radius decreases the ratio between the dislocation core and the remaining NW volume increases. In addition, as the wire surface approaches the core, the displacement field interacts with the wire surface. In these terms, the surface relaxation results in a complex interplay with the screw dislocation properties. For example, if the NW surface shear modulus is smaller than that of the bulk (as predicted for Si NW below 4 nm in diameter [28]), a larger distortion would be needed to compensate the torque. Concerning the precise atomic arrangement at the screw dislocation core, we must keep in mind that the stress at the core diverges as $1/r$; therefore, the perfect helicoid assumed structure must yet be confirmed by additional studies.

It may be interesting to note that the observed large twist generates an additional axial periodicity, which has a scale similar to the exciton Bohr radius for typical III-V semiconductors. We speculate that this new periodicity may influence electrical and optical properties. By analogy, geometrical attributes of carbon nanotubes (diameter, helicity) define transport properties [25]. From a different perspective, the strain fields of the screw dislocation and of the twist create a unique stress profile ([20], Fig. S8). As the gap of a semiconductor depends on the applied stress, the observed structure forms a cylindrical potential well where carriers may be trapped away from the surface and the dislocation core, possibly increasing carriers' lifetimes. Moreover, the size of this tube-like potential well is of the same order as typical semiconductor length scales, possibly leading to quantum interference effects.

To conclude, we have presented a detailed analysis of atomic structure distortions associated with the lattice twisting in wurtzite InP NWs containing a screw dislocation. For NWs with radii in the 10–20 nm radii range, a twist much larger (up to 100%) than estimated from elasticity theory has been measured. This enhancement arises due to the complex interaction between the dislocation displacement field and the free surface. These experiments also illustrate the importance of electron diffraction to characterize deformations in nanometric structures. Moreover, they demonstrate that spatially resolved CBED with 1 nm electron beam experiments can offer valuable information on material behavior and structural distortions.

We thank Paulo Silva for his assistance and the LNNano for equipments used. This work was funded by FAPESP, CNPq, CNRS, IUF, and ESTEEM.

*ltizei@ifi.unicamp.br
†dmugarte@ifi.unicamp.br